\documentclass{article}

\usepackage{arxiv}

\usepackage{tikz}
\usepackage{amsmath}

\usepackage{url}
\usepackage{enumerate}   
\usepackage{amssymb}

\usepackage{threeparttable}
\usepackage{diagbox}

\usepackage{graphicx}
\usepackage{float}

\usepackage{inconsolata}

\usepackage{multirow}

\usepackage{textcomp}
\usepackage{xcolor}
\usepackage[utf8]{inputenc}
\usepackage{listings}
\usepackage[draft]{minted}
\usepackage{listings}

\usepackage{filecontents}
\usepackage{nameref}
\usepackage{xspace}
\usepackage{paralist}
\usepackage{enumitem}
\usepackage{pifont}
\usepackage{listings}
\usepackage[affil-it]{authblk}

\usepackage[english]{babel}
\usepackage{blindtext}

\newcommand{\tab}{\hspace*{1em}}
\newcommand{\code}[1]{{\fontfamily{cmtt}\fontseries{m}\fontshape{n}\selectfont\small{#1}}}

\newcommand{\sysname}{SGXCapsule\xspace}

\begin{document}

\date{}

\title{Towards Efficiently Establishing Mutual Distrust Between Host Application and Enclave for SGX}

\newcommand\CoAuthorMark{\footnotemark[\arabic{footnote}]}

\author[1]{Yuan Chen}
\author[1]{Jiaqi Li}
\author[1]{Guorui Xu}
\author[1]{Yajin Zhou\thanks{Corresponding author (yajin\_zhou@zju.edu.cn).}\hspace*{0.4em}}
\author[2]{Zhi Wang}
\author[3]{Cong Wang}
\author[1]{Kui Ren}
\affil[1]{Zhejiang University}
\affil[2]{Florida State University}
\affil[3]{City University of Hong Kong}

\maketitle

\begin{abstract}
	Since its debut, SGX has been used in many applications, e.g., secure data
	processing. However, previous systems usually assume a trusted enclave and ignore the
	security issues caused by an \textit{untrusted} enclave.
	For instance, a vulnerable (or even malicious) third-party enclave can be exploited to
	attack the host application and the rest of the system.
	In this paper, we propose an efficient mechanism to
	confine an untrusted enclave's behaviors. The threats of an untrusted enclave
	come from the enclave-host asymmetries. They can be abused to access arbitrary memory regions of
	its host application, jump to any code location after leaving the enclave and forge the stack
	register to manipulate the saved context.
	Our solution breaks such asymmetries and establishes mutual distrust between the
	host application and the enclave. It leverages Intel MPK for efficient memory isolation
	and the x86 single-step debugging mechanism to capture the event when an enclave is exiting. Then it performs the integrity check for the jump target and the stack pointer. 
	We have solved two practical challenges and implemented a prototype system.
	The evaluation with multiple micro-benchmarks and representative
	real-world applications demonstrated the efficiency of our system, with
	less than $4\%$ performance overhead.
	
\end{abstract}

\section{Introduction}
Intel Software Guard eXtension (SGX) is one promising hardware
architecture that provides a Trusted Execution Environment (TEE).
With SGX, the sensitive code and data can be put into a protected memory region,
i.e., the \textit{enclave}, which is hardware-isolated from the rest of the system.
Even system software, e.g., the OS and the hypervisor, is unable to access the content of the enclave.

The strong security and privacy guarantees provided by SGX make it attractive to 
build the infrastructure for confidential cloud computing~\cite{vc3, enclavedb, LightBox, haven,graphene, occlum}.
Since the cloud platform is out of the tenant's control and thus untrusted,
SGX addresses this by removing the cloud provider out of the trusted computing base (TCB).

\smallskip
\noindent
\textbf{Problem statement}\tab
However, previous systems usually assume a trusted enclave and ignore the
security issues caused by an \textit{untrusted} enclave.
Such an assumption is problematic from the perspective of an enclave's host application.
One representative scenario is the third-party enclaves.
With the popularity of SGX, service providers tend to deploy their services
(e.g., trained machine learning models) inside the enclave to protect their intellectual property.
This greatly simplifies the interactions between the service providers and consumers, thus creating valuable business benefits.
However, embedding the service provider's enclave into an application brings huge
security risks to users, considering the service enclave comes from an (untrusted) third-party.
Even worse, the isolation provided by SGX makes it much harder to inspect the third-party enclave.
Recent research~\cite{sgxrop} has demonstrated the hazard of an enclave malware to hijack its host
application's control flow~\textit{stealthily}.
Furthuremore, potential software bugs in an enclave also makes it untrusted to the host application.
For instance, 
Van Bulck et al.~\cite{teeassess} vetted six popular SGX runtimes and found security issues in all of them.
A buggy enclave can be exploited by the attackers to compromise its host application and
even the system software further.

\smallskip
\noindent
\textbf{Current solutions and their limitations}\tab
Because of the isolation provided by SGX, an enclave is prevented from the inspection by
the anti-virus system at runtime~\cite{sgxexplained}.
Besides, the code in an enclave may be deployed in the ciphertext and only
decrypted inside enclave at runtime to protect the code secrecy~\cite{sgxelide}.
This further impedes introducing a full vetting process on the enclave code.

One proposal is to detect the malicious actions of an enclave by monitoring its
I/O behaviors~\cite{sgxexplained}, which is not practical yet.
Moreover, Costan et al. discussed to embed a standardized static analysis framework into
the enclave~\cite{sgxexplained}. This brings two security concerns, i.e., the source credibility
of the static analysis framework and the increased TCB introduced by the framework.
As the first implemented defense system, SGXJail~\cite{sgxjail} leverages the process isolation to confine an untrusted enclave.
However, it requires creating a dedicated sandbox process for each enclave. Thus it is not scalable, especially for multiple enclaves and multithreading.
Weiser et al. proposed a hardware-based defense mechanism, named HSGXJail~\cite{sgxjail}.
HSGXJail requires the hardware modification and is hard to be deployed in practice.

\smallskip
\noindent
\textbf{Our solution}\tab
In this paper, we propose  an efficient defense mechanism to confine an untrusted enclave's behaviors.
Threats caused by an untrusted enclave are due to the
blind trust of the host application to the enclave~\cite{sgxjail}.
Such blind trust causes the enclave-host asymmetries,
i.e., \textit{the data access asymmetry and the control flow asymmetry}.

Specifically, with the data access asymmetry, an enclave can read and write arbitrary memory regions of
its host application, while the memory region of an enclave is hardware-protected by SGX.
With the assistance of Intel \textit{Transactional Synchronization Extensions} (TSX), an enclave can
even probe the whole address space of its host application \textit{stealthily} without triggering any exceptions~\cite{sgxrop}.
Besides, with the control flow asymmetry, an enclave can jump to \textit{any code location of its host application}
after leaving the enclave and forge the stack register to manipulate the saved context.

Our solution intends to break such asymmetries and establishes mutual distrust between the host application and the enclave.
To this end, it leverages two x86 hardware features,
i.e., \textit{Intel memory protection key (MPK) and x86 single-step mode}, to break the data access and control flow asymmetries, respectively.
Specifically, it leverages MPK for efficient memory isolation. It assigns different protection keys for the host
application and the enclave so that the enclave cannot access arbitrary memory regions of its host application. It
leverages a shared buffer for the data sharing crossing the enclave boundary with a same protection key.
Moreover, it leverages the x86 single-step debugging mechanism to capture the event when an enclave is exiting. It then
performs the integrity check for the jump target and the stack pointer.

However, using MPK to confine the data access of an enclave is not trivial. 
Because the MPK's access permission policy can be manipulated with two \textit{user-mode} instructions,
i.e., \code{WRPKRU} and \code{XRSTOR}, inside the enclave. We need to ensure such instructions cannot be
abused by the enclave to bypass our solution.
Moreover, the hardware isolation of SGX makes it even harder when the code secrecy of an
enclave is protected~\cite{sgxelide}, since the code inside the enclave cannot be vetted by the host application.

To prevent the enclave from abusing the \code{XRSTOR} instruction
to change the MPK's access policy, we leverage the hardware configuration of the
ninth bit of the enclave's XFRM attribute filed to disable the update of the policy. If this bit
is \textit{cleared}, the MPK's access policy cannot be updated through the \code{XRSTOR} instruction inside the enclave.
Thus, the host application can ensure this bit has been cleared when creating an enclave.
For the \code{WRPKRU} instruction, our solution chooses to embed a piece of inspection
code inside the enclave and enforces the \textit{W \(\oplus\) X} policy for the enclave memory
during the life-cycle of the enclave to capture the dynamically generated and runtime decrypted code.
This enables our system to guarantee that any code inside the enclave is inspected before being executed.
The inspection code ensures the \code{WRPKRU} instruction is not present in the enclave's code.
Since our solution is based on in-process confinement, it is more scalable compared
to the previous system~\cite{sgxjail}, which relies on the process isolation.
Our solution does not introduce any performance overhead  for the code execution inside the enclave.
The overhead only occurs when the execution crosses the boundary of the host application and the enclave.

We have implemented a prototype system named \sysname based on Intel SGX SDK for Linux.
Note that our design is coupled with neither Intel's SGX SDK nor Linux.
It is applicable to other SGX runtimes and OSs as long as MPK and x86 single-step mode are
supported.

We use multiple micro-benchmarks and real-world applications to evaluate the performance overhead
of our system.
In particular, we use three representative applications, i.e., 
the privacy-preserving machine learning service, the relational database, and a HTTPS web server in the evaluation. 
The machine learning service usually requires the large size parameter passing, e.g., for model weights or inputs.
The database and a web server require lots of system calls, which represents the
scenarios of high-frequent context switches between the host application and the enclave.
The evaluation result shows that our prototype is efficient.
It only introduces an average performance overhead of $0.84\%$ for the machine learning service, $1.26\%$
for the database and $3.98\%$ for the HTTPS web server.

In summary, this paper makes the following main contributions.

\begin{itemize}[nosep,leftmargin=1em,labelwidth=*,align=left]
	\item We summarize two types of the enclave-host asymmetries and work towards establishing mutual distrust between them.
	In particular, we leverage two x86 hardware features, i.e., MPK and the single-step mode, to efficiently break
	the asymmetries (Section~\ref{subsec:dataaccess} and Section~\ref{subsec:controlflow}).
	
	\item We have solved two practical challenges,
	i.e., blocking PKRU update inside the enclave 
	and host stack pointer manipulation by the enclave (Section~\ref{subsec:challenge}),
	and implemented a prototype system named \sysname (Section~\ref{sec:implementation}).
	
	\item The evaluation with multiple micro-benchmarks and representative applications
	shows the efficiency of our system, with
	less than $4\%$ performance overhead (Section~\ref{sec:evaluation}).
\end{itemize}

To engage the community, we will release the source code of our system.

\section{Background}
\subsection {Intel SGX}
\label{subsec:sgx}

Intel Software Guard eXtension (SGX) allows an application to create
a so-called \textit{enclave}, which contains sensitive information (code and data).
The confidentiality and integrity of an enclave are hardware-guaranteed, even system software (e.g., OS and hypervisor) cannot access the content of the enclave.
Architecturally, an enclave is a protected memory region residing in the host application's address space.

\smallskip
\noindent
\textbf{Host-enclave interaction}\tab
The programming model of the Intel SGX SDK~\cite{intelsdk} allows developers to specify two kinds of host interfaces for the enclave.
First, an invocation into the enclave is referred as an \code{ECALL}, which is used by the
host application to invoke a specific pre-defined function inside the enclave.
Second, \code{OCALL}s are used by the enclave to invoke the functions outside,
which are usually aimed to request OS services (e.g., system calls).
Besides, the SDK performs proper sanitizing and marshaling for
the \code{ECALL/OCALLs}' parameters according to the argument attributes specified by the enclave developer.
For instance, for a data pointer argument of an \code{ECALL} with the
\code{[in,size=10]} attribute, the SDK would allocate a
10-byte memory buffer inside the enclave and
copy the corresponding data from the host application into the allocated buffer
inside the enclave.

The \code{ECALL/OCALL} interface functions are defined via a so-called {Enclave Definition Language (EDL)}.
The \code{Edger8r} tool, shipping as part of the SDK, can generate (dispatch/receiving) edge routines fo
\code{ECALLs/OCALLs} according to developer-defined EDL files.
At runtime, these edge routines reside in both the host application and the enclave for routing \code{ECALL/OCALL} requests.

Besides, the {Trusted Runtime System} (tRTS) and {Untrusted Runtime System} (uRTS) are provided
to embed into an enclave and its host application to perform enclave management and \code{ECALL/OCALL} requests routing.
The dispatch (receiving) edge routines resides in both sides would send (accept)
the \code{ECALL/OCALL} requests to (from) the tRTS/uRTS and redirect them from (to) the user code.
The combination of the tRTS/uRTS and the edge routines facilitate
the (host application/enclave) developers to invoke an \code{ECALL/OCALL} as a function call.

The implementation of \code{ECALL/OCALL} interfaces is based on two \textit{user-mode}
instructions, \code{EENTER} and \code{EEXIT}, respectively.
Specifically, the \code{EENTER} instruction is used by the host application to transfer
the control to a pre-defined address inside the enclave.
The \code{EEXIT} instruction allows the execution leave the enclave. 
As an operand of the \code{EEXIT} instruction,
the \code{RBX} register specifies the jump target address
of \code{EEXIT} instruction outside the enclave and will be filled into
\code{RIP} register after the execution leaves the enclave.
Therefore, \textit{the jump target of the \code{EEXIT} instruction is enclave-manipulatable.} 

Moreover, SGX leaves most of the execution states (e.g. registers) switching and sanitizing to the software when the execution crosses the boundary of the host application and the enclave via the \code{EENTER} and \code{EEXIT} instructions.
For instance, when the execution wants to leave the enclave via the \code{EEXIT} instruction,
it is the responsibility of the enclave code to refill the host stack pointers, i.e., \code{RSP} and \code{RBP}.
Thus, \textit{the enclave can refill a fake stack pointer inside the host application to manipulate the saved context.}

Furthermore, the enclave execution can be aborted asynchronously when an exception
(e.g., a page fault or a hardware interrupt) occurs.
This is referred as \textit{asynchronous enclave exit} (AEX).
On an AEX event, the processor would save the enclave's current execution context, such as general purpose registers (GPRs) and processor extended states, into the enclave's \textit{state save area} (SSA) frame before exiting the enclave.
After an AEX event, the host application can reenter the enclave and resume the previously saved execution context of the enclave via the \code{ERESUME} instruction.

SGX supports multithreading with a special enclave data structure, named \textit{Thread Control Structure} (TCS).
A TCS contains information for a thread execution inside an enclave, such as the address of the enclave entry function and the relative address of SSA.
The content of a TCS is specified by the enclave (developers) and is \textit{not accessible} by the software (including the enclave) after being initialized.
Whenever the execution wants to enter the enclave (via \code{EENTER/ERESUME}), a free TCS needs to be specified in the instruction operand.
By doing this, SGX allows concurrent execution inside the enclave.

\subsection {Intel MPK}

Memory Protection Key (MPK) is a new hardware feature introduced in recent Intel processors to provide permission control over the page groups from a per-thread view.
MPK exploits four previously-unused bits of the page table entry to serve as the page's protection key.
Thus, MPK partitions a process's address space into 16 disjoint protection domains.
To enforce the per-thread permission control, a per-core 32-bit \textit{protection key rights register} (PKRU) is introduced.
Every two bits in PKRU determine the access permission to one specific protection key. 
Specifically, PKRU[\textit{2i}] represents the {access-disable} bit while PKRU[\textit{2i+1}] represents {write-disable} bit for protection key \textit{i}.
Based on {access-disable} and {write-disable} bits, three permission policies are enforced: {read/write}, {read-only} and {no-access}.

To access PKRU, MPK introduces two new \textit{user-mode} instructions, \code{RDPKRU} (for reading) and \code{WRPKRU} (for writing).
Updating the value of the PKRU register with the \code{WRPKRU} instruction has negligible overhead, which takes less than 20 cycles~\cite{libmpk}. 
This makes MPK a highly efficient user-space memory permission control primitive.
Note that the {execution permission} is \textit{not} impacted by MPK.

\smallskip
\noindent
\textbf{Interact with SGX}\tab
During the enclave execution, the value of PKRU can be stored into and
restored from memory as part of the \textit{processor's extended states},
as long as the bit 9 in the enclave's XFRM attribute field is set.
Inside the enclave, the processor's extended states can be stored into and restored from memory in two scenarios.
First, an enclave can save and restore them via \code{XSAVE} and \code{XRSTOR} instructions.
Second, the processor's extended states are saved into the enclave's \textit{state save area} (SSA) frame on an AEX event
and restored when reentering the enclave with the \code{ERESUME} instruction.

\subsection {The x86 Single-Step Mode}

The x86 architecture supports the single-step mode for debugging.
It allows the processor to generate a trap after executing each instruction.
The single-step mode is activated by setting a bit, named {Trap Flag} (TF), within the processor's \code{FLAGS} register.
If the TF bit is set by an application using a \code{POPF, POPFD}, or \code{POPFQ} instruction,
a single-step debug exception is generated after the instruction following the corresponding instruction.
Accordingly, the single-step mode can be disabled by clearing the TF bit.

\smallskip
\noindent
\textbf{Interact with SGX}\tab
For the opt-out enclave (i.e. enclave with the debugging feature disabled), 
if the TF bit is set at the time of the enclave entry, a single-step debug exception would be pending immediately after exiting the enclave via the \code{EEXIT} instruction.
In other words, the processor in the single-step mode treats the whole life-cycle of the enclave execution (from enclave entry to enclave exiting) as a single instruction.
Besides, the enclave is \textit{not allowed} to manipulate the value of the TF bit.
The processor guarantees this with two properties.
First, at the time of the enclave entry, the processor stores the TF bit into a software \textit{invisible} register and then clears it, while the value of the TF bit is restored after exiting the enclave.
Second, the processor ensures that the TF bit always remains cleared inside the enclave.
Note that if there is an AEX event happens inside the enclave, the processor will
exit the enclave without pending a single-step debug exception.

\section{Threat Model}
Our threat model considers the mutual distrust between the
enclave and its host application. Therefore, the threat model consists of
the following
two perspectives.

\begin{itemize}[nosep,leftmargin=1em,labelwidth=*,align=left]
	\item \textbf{From the enclave's view}\tab 
	It assumes the same attack model as other systems that leverage SGX for protection.
	Specifically, the trust anchors for an enclave (developer) are the SGX-enabled CPU and the code inside the enclave.
	The rest of the software stack (including the host application, OS, and hypervisor) is considered as untrusted.
	The goal of an enclave is to prevent the leakage of its sensitive information (data or/and code) both at runtime and at rest.
	Thus, code secrecy mechanism~\cite{sgxelide} may be used by an enclave to protect its private code.
	
	\item \textbf{From the host application's view}\tab
	It considers the enclave as~\textit{untrusted}~\cite{sgxrop,sgxjail}.
	As mentioned previously, such scenarios include third-party enclaves and potentially buggy enclaves.
	The goal of the host application is to leverage the functionalities provided by the enclave
	while confining the enclave's behaviors.
\end{itemize}

\smallskip
\noindent
\textbf{Out of scope}\tab
We only consider restricting the host-enclave interactions to specified interfaces.
The high-level attacks, such as the Iago attack~\cite{iago} that exploit the specified interfaces are ignored by us.
It is because that we regard our work as the first step towards \textit{the establishment of the mutual distrust between the enclave and its host application}.
Besides, side-channel and denial-of-service (DoS) attacks are not considered.
Defending against such attacks are orthogonal to our work.

\section{System Design}

From the enclave's view, the security mechanisms provided by SGX can be used to protect the sensitive data and code inside the enclave. However, from the host application's view, there exist two types of asymmetries~\cite{sgxjail,sgxrop} between the enclave  and its host application.

\begin{itemize}[nosep,leftmargin=1em,labelwidth=*,align=left]
	
	\item \textbf{Data access asymmetry}\tab
	An enclave can read and write arbitrary memory regions of its host application, while the memory region of an enclave is hardware-protected by SGX.
	With the assistance of Intel {Transactional Synchronization Extensions} (TSX), an enclave can even probe the whole address space of its host application \textit{stealthily} without triggering any   exceptions~\cite{sgxrop}.
	\item \textbf{Control flow asymmetry}\tab
	An enclave can jump to any code location of its host application by specifying the target address
	inside the \code{RBX} register, which will be used by the \code{EEXIT} instruction. It can also forge
	the stack register when leaving from the enclave to manipulate the saved context (Section~\ref{subsec:sgx}).
	However, the host application is not allowed to specify the entry point when entering the
	enclave via the \code{EENTER/ERESUME} instruction.
\end{itemize}

\smallskip
The main purpose of our work is to confine the \textit{untrusted}
enclave for its host application, so that the host-enclave interaction can be restricted to the specified interfaces.
This helps to establish the mutual distrust between them.

To achieve this goal, \sysname leverages two x86 hardware features to efficiently eliminate the enclave-host asymmetries.
First, for data access asymmetry, \sysname relies on Intel MPK~\cite{libmpk} to efficiently limit the enclave access to limited regions of the host memory (Section \ref{subsec:dataaccess}).
Only some specific regions in the host memory's address space are accessible to the enclave.
These regions serve as parameter passing buffers for the host-enclave interaction.
Second, for control flow asymmetry, \sysname   leverages the x86 single-step mode to ensure that the execution continues from the pre-defined location after leaving the enclave (Section \ref{subsec:controlflow}).

\begin{figure}[t]
	\centering
	\includegraphics[width = 0.6\linewidth]{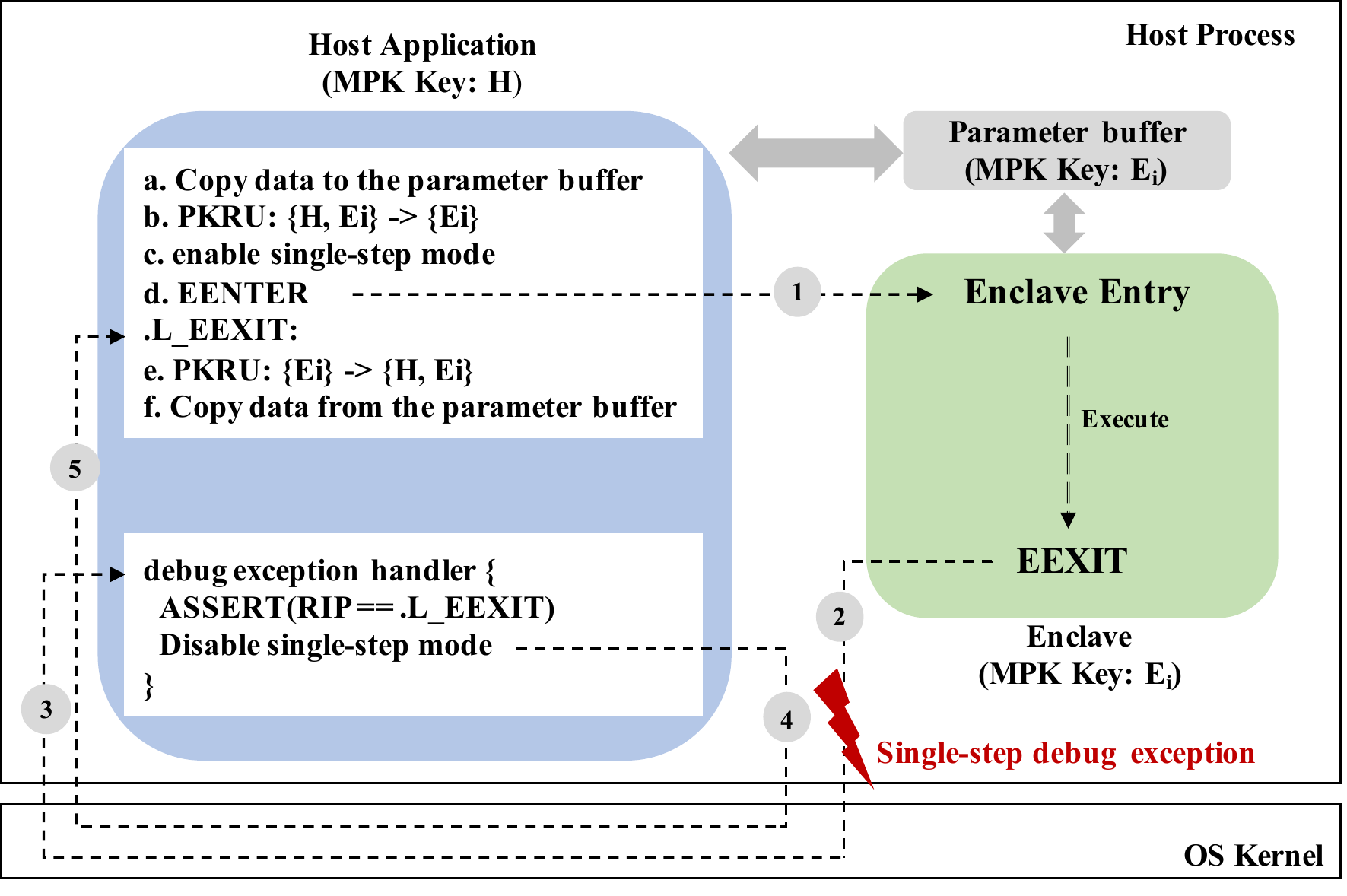}
	\caption{The over system architecture.}
	\label{fig:design}
	\vspace{-0.2in}
\end{figure}

It is not trivial to enforce such confinements. There are two challenges that need to be solved, i.e., how to block the update the PKRU register inside the enclave and how to protect the host application's saved context when leaving the enclave. We will illustrate our solutions in Section~\ref{subsubsec:c1} and Section~\ref{subsubsec:c2}, respectively.
Figure \ref{fig:design} shows the overall system architecture.

\subsection{Data Access Asymmetry Elimination}
\label{subsec:dataaccess}

To eliminate the data access asymmetry, MPK is used to restrict the data access of an enclave
to the specified regions in its host application's address space.
For simplicity, we first assume there is only one enclave \textit{i} in the host
application and then discuss the scenario of multiple enclaves.

The host memory regions are assigned with the protection key
\textit{H}, which is exactly the default protection key in the OS (e.g., key 0 in Linux).
\sysname then assigns the memory region of enclave \textit{i} with a unique protection key \textit{Ei} at the time of the enclave creation.
To support parameters passing for the host-enclave interaction, \sysname couples each TCS of the enclave \textit{i} with a piece of the host memory region, called the \textit{parameter buffer}.
A parameter buffer is organized as a stack to enable the \textit{nested} host-enclave interactions.
Because a parameter buffer is coupled with each TCS, \sysname can support concurrency inside the enclave without the \textit{scalability} issue.
Accordingly, All the parameter buffers  of the enclave \textit{i} belongs to the MPK protection key \textit{Ei}.

First, the host application has the access permission to protection keys \{\textit{H}, \textit{Ei}\}.
This does no harm to the security of the enclave since the content of the enclave is hardware-protected by SGX.
Second, the enclave can only access the protection key \textit{Ei}.
Before the execution is transferred to the enclave \textit{i}, the access permission of the current thread is restricted from \{\textit{H}, \textit{Ei}\} to \textit{Ei} via updating the {PKRU} register (step {b} in Figure \ref{fig:design}).
\sysname guarantees that the PKRU register is not changed while the execution remains inside the enclave.
By doing this, the enclave execution cannot access the host memory regions other than the enclave's parameter buffer.
After exiting from the enclave, the execution resumes the host application's original access permission (step {e} in Figure \ref{fig:design}). 
In this way, \sysname eliminates enclave-host data access asymmetry \textit{efficiently}, since no performance overhead is introduced inside the enclave.

\smallskip
\noindent
\textbf{Multiple enclaves}\tab
The support of multiple enclaves is straightforward.
First, a unique protection key is allocated for each enclave.
The memory region and the parameter buffer of an enclave are then assigned with the enclave's
protection key during the process of the enclave creation.
Second, the host application has access permission to all the enclaves' protection keys,
whereas each enclave can only access its own protection key. If the number of enclaves
exceeds the maximum number of available protection keys,
the virtualization of protection keys could
be leveraged~\cite{libmpk}.

\subsection{Control Flow Asymetry Elimination}
\label{subsec:controlflow}

In Section~\ref{subsec:dataaccess}, we show how \sysname utilizes MPK to eliminate
the data access asymmetry.
However, the design goal of MPK is intended to provide permission control for data memory access. Instruction fetching is not constrained by MPK. Hence, MPK cannot be used to eliminate control flow asymmetry. Thus, the code inside the enclave can jump to arbitrary executable code locations inside the host application's memory space after executing the \code{EEXIT} instruction.

The root cause of the control flow asymmetry is that the jump target of the \code{EEXIT} instruction is enclave manipulatable.
To solve this problem, \sysname leverages the x86 single-step mode feature to detect whether the jump target of the \code{EEXIT} instruction matches the pre-defined location, i.e., the next instruction after the \code{EENTER} instruction (label .L in Figure~\ref{fig:design}).

Specifically, whenever the execution gets into the enclave, the TF bit within the \code{FLAGS} register is set (step {c} in Figure \ref{fig:design}) to ensure the pending of a single-step debug exception follows the \code{EEXIT} instruction (\ding{173} in Figure~\ref{fig:design}).
The corresponding exception handler then performs the check for the jump target (\ding{174} in Figure \ref{fig:design}).
If the check passes, the execution is resumed and continues from the pre-defined location in the host application's code region (\ding{176} in Figure \ref{fig:design}).
Otherwise, the potential abuse (or exploit) of the control flow asymmetry is detected. The execution is aborted.

Same with the MPK-based data access asymmetry elimination, eliminating control flow asymmetry with the assistance of the single-step mode introduces no performance overhead for the execution \textit{inside the enclave}.

\subsection{Challenges and Solutions}
\label{subsec:challenge}

We have described the main idea of adopting two x86 hardware features to eliminate the enclave-host asymmetries.
However,
there are still some challenges to ensure that the security enforcement of our system cannot be bypassed.
In this following, we illustrate these challenges and our solutions.

\subsubsection{Challenge I: Block PKRU Update Inside the Enclave}
\label{subsubsec:c1}
The PKRU register can be updated with two \textit{user-mode} instructions, \code{WRPKRU} and \code{XRSTOR}.
Note that for \code{XRSTOR}, the \code{PKRU} register can be affected only if bit 9 within the enclave's XFRM attribute field is set.
An untrusted enclave may use these two instructions to extend its access permission to the protection keys other than itself.
To prevent this, \sysname needs to keep the {PKRU} register unchanged during the enclave execution.

\begin{listing}[t]
	\begin{minted}
		[
		linenos,
		numbersep=3pt,
		frame=single,
		fontsize=\scriptsize,
		highlightlines={7-9,16-18},
		]{c}
		/******** wrapper of WRPKRU ********/
		xor  %
		xor  %
		mov  $PKRU_DISALLOW_TRUSTED, %
		WRPKRU             // %
		//the inserted code is as follow
		cmp  $PKRU_DISALLOW_TRUSTED, %
		je   .Lcontinue
		syscall  exit
		.Lcontinue:
		// execution continues...
		
		/******** wrapper of XRSTOR ********/
		XRSTOR
		// the inserted code is as follow
		bt  %
		jnc .Lcontinue
		syscall exit
		.Lcontinue:
		// execution continues...
	\end{minted}
	\caption{The wrappers of WRPKRU/XRSTOR in ERIM}
	\label{list:erim}
	\vspace{-0.1in}
\end{listing}

A similar challenge is also addressed by the ERIM system~\cite{erim}. However, the solution proposed by the ERIM cannot be directly applied to our system.
\begin{itemize}[nosep,leftmargin=0em,labelwidth=*,align=left]
	
	\item[\textbf{CI-1)}] \textit{The register value could be changed after {WRPKRU/XRSTOR} instructions.} ERIM leverages the wrappers for \code{WRPKRU/XRSTOR} instructions for a safe PKRU update. As shown in Listing \ref{list:erim}, a few instructions are inserted after \code{WRPKRU/XRSTOR} instructions to check whether the PKRU register is updated as expected. The assumption is that the value of the \code{EAX} register between the \code{WRPKRU} instruction (line 5) and the check of the \code{EAX} value (line 7) cannot be changed. The same assumption applies to the \code{XRSTOR} (line 14) and the following check routine (line 16). However, this assumption does not hold in our scenario.   
	When the enclave execution is interrupted on an AEX event and is going to exit the enclave, the program states of the execution would be stored into the enclave's SSA frame by hardware.
	The SSA frame resides inside the enclave and is software visible.
	Other enclave executions, if existing, then can corrupt the program states of the interrupted execution by
	modifying the content of the corresponding SSA frame.
	
	\item[\textbf{CI-2)}] \textit{The dynamically loaded code inside the enclave could be encrypted.}
	Besides, the dynamically loaded and runtime-generated code make the 
	problem even harder. Some enclaves may load or generate code at runtime for the purpose of code secrecy~\cite{sgxelide}.
	For instance, an enclave developer may want to protect its proprietary code.
	So he/she chooses to deploy the proprietary code in the ciphertext.
	A small piece of the code loader is included in the enclave to load and
	decrypt the proprietary code after enclave creation and the remote attestation.
	The hardware isolation of SGX prevents the host application's inspection system, such as the interception mechanism proposed in ERIM~\cite{erim}, from inspecting the newly loaded or generated code.
	
	\item[\textbf{CI-3)}] \textit{The occurrence of the XRSTOR instruction is unavoidable inside the enclave.}
	One intuitive thought is to forbid the occurrence of \code{WRPKRU/XRSTOR} instructions inside the enclave.
	However, SGX leaves the switching of the execution's program states to
	the software (i.e., enclave code) when the execution crosses the enclave boundary
	via \code{EENTER/EEXIT} instruction.
	Thus, \code{XSAVE/XRSTOR} instructions are usually used by enclave to efficiently
	save/restore the extended states of the execution.
	The occurrence of the \code{XRSTOR} instruction is unavoidable inside the enclave.
\end{itemize}

The goal is keeping the PKRU register remain unchanged during the enclave execution.
To address the above challenges, \sysname adopts different strategies for the
\code{XRSTOR} and the \code{WRPKRU} instruction, respectively.

\smallskip
\noindent \textbf{XRSTOR instruction}\tab
The usage of the \code{XSAVE/XRSTOR} instruction
is intended to save/restore the current execution's extended states \textit{efficiently}.
However, in our design, an enclave is not allowed to maintain the PKRU register as part of the enclave execution's extended states.
Because \sysname requires the PKRU register keep unchanged during the enclave execution.
Besides, there is no need for an enclave to maintain its own PKRU register value.
That's because \textit{MPK cannot be used by an enclave to provide access permission control for itself}.
Because the permission control provided by MPK is based on two components: the per-core PKRU register and the 4-bit protection key field in the page table entry.
Apparently, the latter one is out of the enclave's control.
Thus, from the enclave's perspective, if it leverages MPK to enforce the permission control, the untrusted host application or operating system can easily corrupt it.

We observed that the PKRU register can be updated by the \code{XRSTOR} instruction inside the enclave \textit{only when the
	bit 9 within an enclave's XFRM attribute field is set.}
Therefore, our solution is straightforward. We require that the bit 9 within an enclave's XFRM attribute field is never set.
Specifically, during the enclave creation, \sysname performs an additional check.
If the bit 9 within the enclave's XFRM attribute field is set, \sysname refuses to create the enclave.
Since the XFRM attribute field cannot be modified after the enclave creation, the created enclave can never use the \code{XRSTOR} instruction to manipulate the PKRU register. With this method, \sysname guarantees that the \code{XRSTOR} instruction cannot affect the PKRU register even though it occurs inside the enclave (Challenge I-3).

\begin{figure}[t]
	\centering
	\includegraphics[width = 8.1cm]{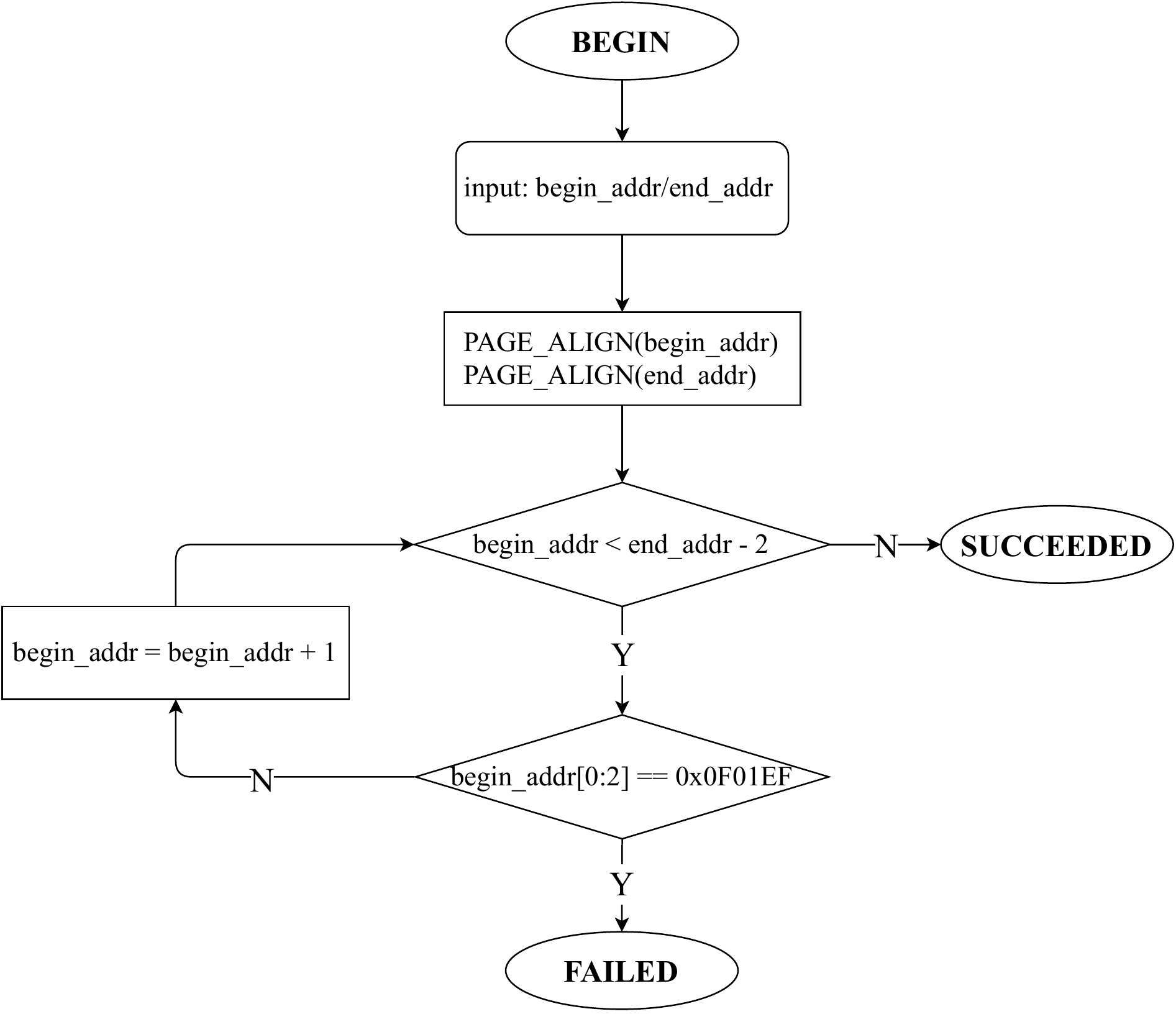}
	\caption{The flow chart of the embedded inspection code.}
	\label{fig:inspect}
\end{figure}

\smallskip
\noindent \textbf{WRPKRU instruction}\tab
\sysname prevents the occurrence of the \code{WRPKRU} instruction inside the enclave.
Compared to ERIM, this simplifies the design, since we no longer need to consider the design of WRPKRU wrappers (Challenge I-1).
To this end, a binary inspection mechanism is needed.
Such an inspection mechanism should be able to handle the dynamic loaded or generated code (abbreviated as \textit{DLGC}) inside the enclave (Challenge I-2).

The host application can detect the execution of DLGC by enforcing the \textit{W \(\oplus\) X}
attribute for the enclave's page table entries.
Two possible options can be made: forbidding DLGC executing inside the enclave 
or requiring an enclave to submit the DLGC to its host application voluntarily to get execution permission.
However, neither of them work in practice.
The former limits the functionalities of an enclave, while the latter brings intellectual property issues and is impossible to verify the authenticity of the submitted DLGC.

Instead, \sysname chooses to \textit{embed} a piece of inspection code in the enclave.
The inspection code is invoked when the host application needs to decide whether to give execution permission to the DLGC.
Our design is based on the observation that the task of our inspection code is just scanning the corresponding code region linearly to detect the occurrence of a specific byte sequence (i.e., \code{WRPKRU}'s machine code: \code{0x0F01EF}).

As shown in Figure \ref{fig:inspect}, the logic of our inspection code is quite simple and can be integrated into the plain code of an enclave without any security concern.
At the time of the enclave creation, \sysname would check the inspection code is indeed included within the enclave's plain code.
\sysname then performs a simple control flow analysis, starting from the enclave entry, to ensure the inspection code can be invoked by the host application. These two ensure that our inspection code exists and is executable.
To simplify the control flow analysis, \sysname places the inspection code as near to the enclave entry as possible.
Note that we can see from Figure \ref{fig:inspect}, the embedded inspection code supports binary inspection at the page granularity and only uses registers to store its data.

We have described that \sysname  can enforce no occurrences of the \code{WRPKRU} instruction
in the DLGC of an enclave with two components:
\textit{W \(\oplus\) X}   for the detection of DLGC and the embedded code for inspection.
In the following, we illustrate the overall workflow of the binary inspection mechanism.
\begin{itemize}[nosep,leftmargin=1em,labelwidth=*,align=left]
	\item \textbf{Static binary inspection}\tab
	This mechanism targets for the plain enclave code.
	At the time of the enclave creation, \sysname scans the loaded plain enclave code for inspection.
	If the inspection passes, the corresponding enclave code page would be assigned with the execution permission.
	Note that during the process of the enclave creation, \textit{W \(\oplus\) X} is always enforced for all the enclave pages.
	With the static binary inspection, \sysname achieves two guarantees.
	First, there is no \code{WRPKRU} instruction inside the enclave code after the enclave initialization.
	Second, the execution of DLGC inside the enclave would be detected by \sysname.
	\item \textbf{Dynamic binary inspection}\tab
	There are two ways to invoke the dynamic binary inspection.
	First, an enclave requests the dynamic binary inspection on its own initiative.
	This indicates that the enclave code is aware of the existence of the dynamic binary inspection. Second, the dynamic binary inspection is transparent to an enclave and triggered by the violation of \textit{W \(\oplus\) X}.

	For both ways, they share the same code inspection process.
	Specifically, \sysname changes the permission of the memory pages that need to be inspected to read-only. This prevents the Time-of-check to Time-of-use (TOCTTOU) attack since the concurrent enclave executions may modify the page during the inspection process.
	Before \sysname specifies a free TCS and gets into the enclave via the \code{EENTER} instruction, it modifies the protection key of the SSA frame of the TCS to the host application's protection key (i.e. \textit{H}).
	This is because if there is an AEX event that occurs during the execution of the embedded inspection code, the register values of the inspection code would be stored into the corresponding SSA frame and may be modified by other concurrent enclave executions.
	Assigning the corresponding SSA frame with the host application's protection key solves this problem.
	Accordingly, \sysname no longer needs to restrict the access permission of the current execution to the enclave's protection key only.

	After that, \sysname gets into the enclave to invoke the embedded inspection code.
	The inspection code would scan the corresponding page(s) according to the address information provided by \sysname.
	One bit is returned by the inspection code to indicate whether the inspection succeeds.
	If the inspection fails, \sysname forbids the execution of the DLGC inside the enclave.
	Otherwise, \sysname assigns the corresponding page(s) with the execution permission. 
	Of course, \textit{W \(\oplus\) X} is still enforced.
\end{itemize}

\subsubsection{Challenge II: Block Host Stack Pointer Manipulation}
\label{subsubsec:c2}
Even though the jump target of the \code{EEXIT} instruction is constrained with the help of x86 single-step mode,
an untrusted enclave is still capable of controlling the host application's stack pointers when the execution gets into the enclave.
In addition, the parameter buffer of an enclave resides outside the enclave and is accessible to the host application.
Thus, a fake host stack can be forged by an untrusted enclave~\cite{sgxrop}.
After the execution leaves the enclave, the program states of the host application are corrupted.

\smallskip
\noindent
\textbf{Our solution}\tab
\sysname borrows the idea from keyed-hash message authentication code (HMAC) to guarantee the integrity of the host stack pointers.
Specifically, \sysname maintains a global 64-bit secret cryptographic key, named \textit{g\_sp\_key}, for the host application.
Whenever the execution gets into the enclave via the \code{EENTER} instruction, \sysname performs bitwise exclusive or (XOR) operations on the host stack pointers,
i.e., \code{RSP} and \code{RBP}, with \textit{g\_sp\_key}, respectively.
The XORed results, \textit{(RSP\(\oplus\)g\_sp\_key)} and \textit{(RBP\(\oplus\)g\_sp\_key)} are then stored into the host stack.
Note that this step takes place before trimming the access permission of the current execution to the enclave's protection key with the \code{WRPKRU} instruction.
Therefore, the execution can still access the host stack.

Since the jump target of the \code{EEXIT} instruction is constrained with the help of the x86 single-step mode, \sysname is able to perform the integrity check for the host stack pointer when the enclave execution finishes.
In detail, \sysname uses the current (\textit{potentially corrupted}) \code{RSP} register for addressing to retrieve the previously stored \textit{(RSP\(\oplus\)g\_sp\_key)} and \textit{(RBP\(\oplus\)g\_sp\_key)}.
Then, \sysname decrypts the retried values with \textit{g\_sp\_key} via XOR operation and compares them with the current \code{RSP/RSP} register.
If their values match, \sysname confirms that the integrity of the host stack pointer is preserved.
Otherwise, the potentially malicious behavior is detected.

Besides, the execution could exit the enclave on an AEX event and the exception handling may need the userspace involvement (e.g., a signal handler in Linux).
For this situation, the integrity check could be performed in the corresponding userspace handler, too.

In summary, \sysname can ensure the integrity for the host stack pointers during the enclave execution due to the following two reasons. First, \textit{(RSP\(\oplus\)g\_sp\_key)}, \textit{(RBP\(\oplus\)g\_sp\_key)} and \textit{g\_sp\_key} are all stored in the enclave invisible memory region.
This is enforced by \sysname with MPK.
Thus, an untrusted enclave is unable to steal or tamper with these values directly. Second, the only information known to the enclave is the plain host
stack pointers. Both \textit{g\_sp\_key} and the result of XOR operation are unknown.
Therefore, the untrusted enclave cannot induce the value of \textit{g\_sp\_key} or forge a valid XORed result in a fake host stack.

\section{Implementation Details}
\label{sec:implementation}

We have implemented a prototype of \sysname based on Intel SGX SDK (v2.9.1) for Linux.
The goal of our prototype is making \sysname transparent to enclave developers and keeping backward-compatibility with the legacy enclave code at the source code level as much as possible.
A developer can port a legacy enclave code into our prototype with minor or even no modifications.
In this section, we illustrate implementation details.

\subsection{Data Access Asymmetry Elimination}
In the following, we describe the implementation details to support the data access asymmetry elimination.

\subsubsection{Parameter passing of ECALL/OCALL}
As shown in Listing \ref{list:edl}, SGX's \code{Edger8r} tool will generate a marshaling data structure based on the argument attributes of an \code{ECALL/OCALL} definition.
The tRTS/uRTS only pass a \code{void} pointer, received from a dispatch edge routine.
As a result, when routing the \code{ECALL/OCALL} requests to the corresponding receiving edge routine, the marshaling data structure is used by the edge routines of both sides to interpret the arguments by converting the \texttt{void} pointer to the marshaling data structure pointer.
By doing so, the tRTS/uRTS do not need to understand the meaning of the \code{void} pointer.
Besides, the \texttt{void} pointer is always pointing to the host application's memory region.
This design assumes an enclave can access its host application's memory regions arbitrarily. This assumption does not hold after applying our system.

\sysname only allows an enclave to access its parameter buffers (\code{param\_buffer}s).
Therefore, we modified the \code{Edger8r} tool and the tRTS/uRTS to support parameter passing of \code{ECALL/OCALL} through \code{param\_buffer}, as shown in the following.

\smallskip\noindent\textbf{Edger8r}\tab
For the \code{Edger8r} tool, we add two wrapper data structures, as shown in Listing \ref{list:edl}, to enable the uRTS to interpret an \code{ECALL/OCALL}'s marshaling data structure.
The dispatch edge routine of an \code{ECALL/OCALL} would instantiate the \code{ms\_param\_meta\_t} and \code{ms\_buff\_meta\_t} to wrap its marshaling data structure.
Then instead of passing a \code{void} pointer, the dispatch edge routine passes the \code{ms\_param\_meta\_t} pointer into the tRTS/uRTS.
The passed \code{ms\_param\_meta\_t} pointer is finally used by the uRTS to copy data to/from the \code{param\_buffer}.

\smallskip\noindent \textbf{tRTS}\tab
The \code{sgx\_ocalloc} function of the tRTS is used by the dispatch edge routine of the \code{OCALL} to allocate memory regions outside the enclave for parameter passing. We modify the implementation of \code{sgx\_ocalloc} to allocate memory from the \code{param\_buffer} instead of the host stack.

\smallskip\noindent
\textbf{uRTS}\tab
The uRTS is responsible for maintaining a \code{param\_buffer} for each TCS of an enclave.
When receiving an \code{ECALL} request, the uRTS chooses a free
TCS and copies the parameters to the corresponding \code{param\_buffer} by interpreting the \code{ms\_param\_meta\_t} pointer received from the dispatch edge routine of the \code{ECALL}.
Besides, an instance of the marshaling data structure of the \code{ECALL} instruction is created in the \code{param\_buffer}, which includes the information about the parameters residing in the \code{param\_buffer}.
Then the uRTS routes the \code{ECALL} request to the tRTS with the address of the newly created marshaling data structure instance as a \code{void} pointer.

When the uRTS receives an \code{OCALL} request from the tRTS, a \code{ms\_param\_meta\_t} pointer that points to some area within the \code{param\_buffer} is also received.
Then the uRTS copies the parameters from the \code{param\_buffer} to a private memory region by interpreting the \code{ms\_param\_meta\_t} pointer.
Besides, an instance of marshaling data structure of the \code{OCALL} is created in the private memory region, which includes the information about the parameters residing in the private memory region.
Note that different from a \code{param\_buffer}, the private memory region belongs to the host application's protection key and thus cannot be accessed by the enclave.
The uRTS maintains a private memory for each TCS of an enclave to prevent the possible TOCTTOU attack on the \code{OCALL} parameters.

\begin{listing}[t]
	\begin{minted}
		[
		linenos, 
		numbersep=5pt, 
		frame=single, 
		fontsize=\scriptsize,
		highlightlines={25-26},
		]
		{C}
		// ECALL definition
		public void ecall_pointer_in_size(
		[in, size=len] void *ptr, size_t len);
		
		
		// generated by Edger8r
		// marshaling data structure for arguments
		typedef struct ms_ecall_pointer_in_size_t {
			void* ms_ptr;
			size_t ms_len;
		} ms_ecall_pointer_in_size_t;
		
		// generated by our modified Edger8r
		// wrapper data structure of ECALL/OCALL's 
		// marshaling data structure
		typedef struct ms_buf_meta_t {
			size_t offset;
			size_t size;
			int in_out;
		} ms_buf_meta_t;
		
		typedef struct ms_param_meta_t {
			void*   ms;
			size_t size;
			ms_buf_meta_t* arr;
			size_t arr_size;
			size_t ret_offset;
			size_t ret_size;
		} param_meta_t;
	\end{minted}
	\caption{
		The \code{ms\_ecall\_pointer\_in\_size\_t} is the marshaling data structure generated
		by \code{Edger8r} according to the argument attributes of the \code{ECALL} definition (Line 3).
		The \code{ms\_buf\_meta\_t} and \code{ms\_param\_meta\_t} are wrapper data structures of \code{ECALL/OCALL}'s marshaling data structure to enable the uRTS to copy the parameters from/to the \code{param\_buf}.
		Specifically, \code{ms\_buf\_meta\_t} is used to describe the pointer member of the marshaling structure (e.g., Line 9).
		\code{ms\_param\_meta\_t} provides the overall description of the marshaling structure.
	}
	\label{list:edl}
	\vspace{-0.1in}
\end{listing}

\smallskip
\noindent
\textbf{Compatibility}\tab
However, our modification to the SDK will disable the support
of the \code{user\_check} attribute.
Specifically, in the Intel SGX SDK, the pointer argument of an \code{ECALL/OCALL}
can be marked using the \code{user\_check} attribute.
For such a pointer, the receiving edge routine will neither
verify the pointer nor copy the pointed buffer.
The design of the \code{user\_check} attribute leaves the duty of
data validation to the developer and is prone to the TOCTTOU attack.
\sysname only supports parameter passing via the developer-invisible
\code{param\_buffers}. Thus, the \code{user\_check} attribute is not
supported in our prototype.

With the implementation above, our prototype keeps source code level backward-compatibility with minor or even no modifications.
The only possible minor modification is to remove the use of the insecure \code{user\_check} attribute.

\subsubsection{Kernel Modifications}

\noindent
\textbf{Issues of the signal handler}\tab
Whenever invoking a signal handler, the Linux kernel will override the
access permission of the current thread to its initial setting (only access to
the protection key zero).
This indicates the kernel assumes the (userspace) signal handler stack belongs to
the protection key zero.
However, during invoking a signal handler, the kernel would access the signal handler's stack (for frame setup or something else) with the thread's original access permission (i.e. the PKRU register value).
This brings up the functional issue when the thread has no permission to the protection key zero.
It can cause a permission violation and terminate the program, since the kernel
accesses the signal handler stack (protection key zero) with no corresponding access permission.

In our prototype, this can happen in two cases. First, the enclave execution exits on an AEX event, and a signal handler is then invoked (e.g., dynamic binary inspection).
Second, the execution leaves the enclave via the \code{EEXIT} instruction and a signal-step debug exception is immediately triggered.  The corresponding signal handler is invoked to check the jump target of the \code{EEXIT} instruction.

\smallskip
\noindent
\textbf{Solution}\tab
Therefore, we make minor modifications (no more $20$ lines of code) to
the Linux kernel.
When the kernel needs to access the signal handler stack, it will always be assigned with the access permission to the protection key zero \textit{temporarily}, no matter what its current permission is.
Specifically, we mainly modify two functions. 
\begin{itemize}[nosep,leftmargin=1em,labelwidth=*,align=left]
	\item \code{handle\_signal}\tab
	Before it calls \code{setup\_rt\_frame} to set up the stack frame for the signal handler, the PKRU register is updated to enable access to the protection
	key zero. One task of \code{setup\_rt\_frame} is saving a copy of the current thread's context (including the PKRU register) into the signal handler's stack.
	Therefore, after returning from the \code{setup\_rt\_frame}, \code{handler\_signal} updates the value of the PKRU register in the copy with its actual value.
	
	\item \code{SYSCALL\_DEFINE0(rt\_sigreturn)}\tab
	This is a system call to return from the signal handler and restore the original context. 
	It will call \code{restore\_altstack} after the original context has been restored.
	Therefore, the PKRU register is updated to enable access to the
	protection key zero before invoking the \code{restore\_altstack} function and resumes after returning.
\end{itemize}

\subsection{Control Flow Asymetry Elimination}

To enable the single-step mode before getting into the enclave via
the \code{EENTER} instruction, the TF bit within the \code{FLAGS} register is set  before the \code{EENTER} instruction via the \code{POPFQ} instruction.
After the execution exits the enclave via the \code{EEXIT} instruction, a single-step debug exception is triggered immediately.
In our implementation, a single handler is registered to catch the \code{SIGTRAP} signal.
In the signal handler, the \code{RIP} register of the original context is checked to see whether it matches the pre-defined location (i.e., the instruction following the \code{EENTER} instruction).
If it fails, the signal handler will report the behavior and abort the execution.
Otherwise, the TF bit in the \code{FLAGS} register of the original context is cleared. Then the execution is resumed from the pre-defined location.

\subsection{Binary Inspection}
To support the code secrecy, Intel SGX SDK provides
\textit{Protected Code Loader} (PCL) mechanism.
With the PCL mechanism, the user enclave code is encrypted at build time
and decrypted at runtime.
To support this mechanism, a static library \code{libsgx\_pcl.a} is included in the
enclave and invoked by the uRTS to decrypt user code immediately
after the enclave creation.
Our prototype provides support for the PCL mechanism and finally implements a three-stage binary inspection mechanism.

\smallskip
\noindent
\textbf{Static binary inspection}\tab
We modify the uRTS to enable it to inspect the \textit{plain} enclave code during the enclave creation.
The modified uRTS also checks that our inspection code is integrated into the enclave and is reachable from the enclave entry.
The uRTS is also modified to enforce \textit{W \(\oplus\) X} during the enclave creation. Specifically, if an enclave code page has the \textit{RWX} permission, the uRTS would restrict its permission to \textit{RW}.

\smallskip
\noindent
\textbf{PCL binary inspection}\tab
Since our prototype enforces \textit{W \(\oplus\) X} inside the enclave, the encrypted user enclave code has only the \textit{RW} permission after the enclave creation.
Our modified uRTS will invoke the embedded inspection code to inspect the newly decrypted user enclave code after the enclave creation.
Note that at this time, there is no concurrency execution inside the enclave, thus no need to change the corresponding enclave page(s) to read-only
or modify the protection key of the corresponding SSA frame to the host protection key.
If the inspection passes, the modified uRTS would change the permission of the enclave page(s) to \textit{RX} via the \code{pkey\_mprotect} system call.

\smallskip
\noindent
\textbf{Runtime binary inspection}\tab
Our prototype relies on the page fault exception to implement the runtime binary inspection.
Since our prototype enforces \textit{W \(\oplus\) X} during the enclave creation, executing newly loaded or generated code inside the enclave would cause a \textit{W \(\oplus\) X} violation and trigger a page fault exception.
A signal handler is registered to catch the \texttt{SIGSEGV} signal.
If the signal code is \code{SEGV\_ACCERR} and the fault page's original permission is \textit{RWX}, the runtime binary inspection is triggered.
The fault enclave page is marked as read-only via the \code{pkey\_mprotect} system call.
The embedded inspection code is then invoked to inspect the fault page.
If the inspection passes, the permission of the fault page is changed to \textit{RX}.
Finally, the original execution is resumed and gets into the enclave via
the \code{ERESUME} instruction.

\section{Performance Evaluation}
\label{sec:evaluation}
In this section, we focus on the performance evaluation of our prototype to demonstrate the effectiveness of \sysname.
Our evaluation was performed on a platform with the Intel i7-10700F CPU (2.90GHz, supports SGX and MPK) and 16GB physical memory.
The system software running on the platform is Ubuntu 18.04.4 (Kernel v5.4.28) with SGX driver v2.6 installed.

\subsection{Evaluation Methodology}

The performance overhead introduced by \sysname is the additional host-enclave interaction latency.
Our \textit{micro-benchmarks} focus on measuring the introduced host-enclave interaction latency for our prototype.
Specifically, we first evaluate the \textit{raw latency overhead}. This is introduced by the additional interaction code, e.g., PKRU register update, single-step mode enabling/disabling, and single-step debug exception handling.
We also evaluate the performance overhead for the \textit{parameter passing}, i.e., copying to/from the parameter buffer (\textit{param\_buffer}).

Besides, in order to see how our prototype works with real-world applications, we choose three representative scenarios, i.e., the privacy-preserving machine learning service, the relational database, and the web server for evaluation.
The choice is based on three reasons:
First, they are popular scenarios for SGX that are evaluated in previous systems.
Second, the database and a web server require high-frequency system calls,
which represent the real scenarios of high-frequent \code{OCALLs}.
Third, as a data-intensive service, the machine learning service usually requires the large size
parameter passing (e.g., for model weights or inputs).

\subsection {Micro-Benchmarks}

\smallskip
\noindent
\textbf{Raw latency overhead}\tab
We implement an empty \code{ECALL} routine without any arguments~\footnote{The \textit{empty} routine means that the worker function of an {ECALL/OCALL} does nothing and returns immediately when it is invoked.}.
We then invoke the empty \code{ECALL} routine from the host application and record the execution time
to represent the raw \code{ECALL} latency.
To measure the raw \code{OCALL} latency, we implement an
empty \code{OCALL} routine without any arguments and invoke it within the
empty \code{ECALL}. Then we record the execution time of invoking the
\code{ECALL} routine again and subtract the raw \code{ECALL} latency.
Last, to evaluate the latency overhead introduced by the data access asymmetry
elimination and the control flow asymmetry elimination, we disable the single-step mode
and perform the above evaluation again.

\begin{table}[t]
	\caption{The evaluation result of the raw ECALL/OCALL latency.}
	
	\begin{center}
		\resizebox{0.5\linewidth}{!}{%
			\begin{threeparttable}
				\begin{tabular}{|c|c|c|c|}
					\hline
					&\multicolumn{1}{c|}{Original}&\multicolumn{1}{c|}{\sysname}&\multicolumn{1}{c|}{\sysname*\tnote{2}} \\
					\hline
					ECALL & $7,636$  & $11,662$ ($52.7\%$) & $9,288$ ($21.6\%$)\\
					\hline
					OCALL & $5,908$ & $9,588$ ($62.3\%$) & $7,303$ ($23.6\%$)\\
					\hline
				\end{tabular}
				
				\begin{tablenotes}
					\footnotesize
					\item[1] Time is measured in CPU cycles.
					\item[2] \sysname* represents our prototype with single-step mode disabled. 
				\end{tablenotes}
			\end{threeparttable}
		}
		\label{eval:latency}
	\end{center}
\end{table}

Table \ref{eval:latency} shows the result.
Overall, our prototype introduces $52.7\%$ overhead for the raw \code{ECALL} latency and $62.3\%$ for the raw \code{OCALL} latency.
The control flow asymmetry elimination contributes $31.1\%$ and $38.7\%$ of the result, respectively.

However, in real-world applications, there are usually complex workloads inside the worker function of the \code{ECALL/OCALL}.
Thus the actual performance overhead of an \code{ECALL/OCALL} introduced by our prototype would be much lower.
In fact, the latency represents the \textit{upper bound} of the performance overhead introduced by our prototype.

\begin{table}[t]
	\caption{The overhead under different execution time inside the enclave and the
		OCALL frequency.}
	\begin{center}
		\resizebox{0.6\linewidth}{!}{%
			\begin{threeparttable}
				\begin{tabular}{|l|c|c|c|c|c|c|c|}
					\hline
					\multirow{2}{*}{Frequency} & \multicolumn{7}{c|}{Execution Time}  \\ \cline{2-8} 
					& 1ms & 5ms & 10ms & 50ms & 100ms & 500ms & 1000ms \\ \hline
					\hline
					1 & 0.2\% & 0.1\% & 0.1\% & 0.04\% & 0.05\% & 0.03\% & 0.03\% \\
					\hline
					10 & 1.4\% & 0.4\% & 0.5\% & 0.06\% & 0.02\% & 0.0\% & 0.0\% \\
					\hline
					100 & 10.8\% & 2.4\% & 1.2\% & 0.3\% & 0.1\% & 0.05\% & 0.03\% \\
					\hline
					1,000 & 42.6\% & 18.0\% & 10.3\% & 2.4\% & 1.2\% & 0.3\% & 0.1\% \\
					\hline
					10,000 & 61.1\% & 51.6\% & 41.5\% & 17.6\% & 10.5\% & 2.4\% & 1.2\% \\
					\hline
					50,000 & 63.7\% & 61.2\% & 56.7\% & 41.6\% & 31.6\% & 10.5\% & \textbf{5.6\%} \\
					\hline
					100,000 & 63.9\% & 62.8\% & 59.8\% & 50.2\% & 42.5\% & 18.0\% & 10.3\% \\
					\hline
				\end{tabular}
			\end{threeparttable}
		}
		\label{eval:demo}
	\end{center}
\end{table}

To further show the raw latency overhead in practice, we use different worker functions that are with
different execution time inside the enclave and various numbers of
\code{OCALL}s invoked during this process. This can represent different types of applications.
Table \ref{eval:demo} shows the result. The execution time inside the enclave ranges from \textit{1ms} to \textit{1s}.
For each test, the function inside the enclave invokes an empty \code{OCALL} without any arguments with different frequencies. We repeat the measurement $20$ times and report the arithmetic mean values.

The overhead is negatively correlated with the execution time inside the enclave, while positively correlated with the \code{OCALL} frequency.
Note that in our evaluation, the \code{OCALL} is empty. Thus the overhead represents the corresponding upper bounds in practice.
Generally speaking, the worker function inside the enclave tends to invoke the \code{OCALLs} at low-frequency, since the main purpose for an \code{OCALL} is to perform the system calls (e.g. file operations).

According to Gruss et al.~\cite{kernelisolation}, Netflix had studied the system call rates of their cloud services and found the highest rate is around $50,000$ system calls per second per CPU.
As shown in Table~\ref{eval:demo},
under the setting that the execution time is 1s and the OCALL frequency is $50,000$, the performance overhead is $5.6\%$.
It demonstrates the effectiveness of our system for real-world scenarios.

\smallskip
\noindent
\textbf{Parameter Passing Overhead}\tab
We use the empty \code{ECALL} and \code{OCALL} routines, with a pointer argument \texttt{buf} whose size is 
64MB. We use a large buffer size to make the time of parameter passing dominate the whole
application's execution, thus omitting the raw latency overhead previously discussed.

After that, we specify the direction attribute of the \texttt{buf} as \texttt{[in]}, \texttt{[out]}, or \texttt{[in, out]} and then
record the time cost for each specification, respectively.
The \texttt{[in]}, \texttt{[out]} and \texttt{[in, out]} for an \code{ECALL} means the parameters will be copied into
the enclave, copied from the enclave and copied in both directions.
The meanings of the attributes for the \code{OCALL} are similar.

Table \ref{eval:param} shows the result.
The worst case is for the pointer argument of an \code{OCALL} with the \texttt{[in]} attribute,
whose overhead achieves 26.4\%. 
Besides, the additional parameter passing time
under the \texttt{[in,out]} setting is obviously longer than that under the \texttt{[in]} or \texttt{[out]} setting.
This is because it requires two additional data copy operations, i.e., copy to and from the \textit{param\_buffer},
while the others only require one operation, i.e. copy to or from the \textit{param\_buffer}. 
Again, this evaluation only considers the raw parameters passing overhead, which shows the upper bound
of the overhead.

One interesting finding is that a pointer argument with the \texttt{[out]}
attribute requires much longer parameter passing time compared to others.
Accordingly, the overhead is much lower (3.7\% and 1.9\%, respectively).
That is because the edge routines of \code{ECALL/OCALL} in the enclave side would
call \code{memset} to clear the buffer that is newly allocated for the pointer argument
with the \texttt{[out]} attribute.
The \code{memset} function is specified as \texttt{\_\_attribute\_\_((optimize("O0")))},
thus the compiled \code{memset} function is not optimized and involves lots of memory access operations.

\begin{table}[t]
	\caption{The evaluation result of parameter passing. Time is measured in milliseconds.  
	}
	\vspace{-0.2in}
	\begin{center}
		\resizebox{0.6\linewidth}{!}{%
			\begin{threeparttable}
				\begin{tabular}{|c|c|c|c|c|c|c|}
					\hline
					&\multicolumn{3}{|c|}{ECALL}&\multicolumn{3}{|c|}{OCALL} \\
					\hline
					& \texttt{[in]} & \texttt{[out]} & \texttt{[in,out]} &  \texttt{[in]} & \texttt{[out]} & \texttt{[in,out]} \\
					\hline
					Original & 28.0 & 145.5 & 45.6 & 17.9 & 157.0 & 45.7 \\
					\hline
					\sysname & 32.9 & 150.9 & 55.7 & 22.7 & 161.1 & 54.4 \\
					\hline
					Additional Time & 4.9 & 5.4 & 10.1 & 4.8 & 3.1 & 8.7 \\
					\hline
					Overhead & 17.6\% & 3.7\% & 22.3\% & 26.4\% & 1.9\% & 18.9\% \\
					\hline
				\end{tabular}

		\end{threeparttable}}
		\label{eval:param}
	\end{center}
\end{table}

\subsection {Macro-Benchmark Evaluation}
\noindent
\textbf{Machine learning as a service}\tab
We run a trained neural network model (with around 34MB weights data) inside an enclave.
The ML model is for predicting the responses of the cancer treatment~\cite{idash}.
SGX is used to protect both the model structure and its data (model parameters, inputs, intermediate results, and the output data).
All of them are encrypted outside the enclave and only decrypted inside the enclave at runtime.
To enable the model inference inside the enclave efficiently, multiple optimization methods, such as multithreading, SGX-friendly memory management, and SIMD with Intel AVX2 extension, have been integrated into the implementation to minimize the inference time.
Besides, due to the limited physical memory inside the enclave, the model performs the inference with a carefully chosen batch size to avoid SGX's inefficient page swapping.

In a word, the implementation of the ML model has been highly optimized to shorten its inference time inside the enclave.
Thus, using it for our experiment is representative to demonstrate the effectiveness of our system.
In our experiment, we perform the model inference on $392$ inputs (total size is 19MB) and record
the time cost of each stage (i.e., enclave creation, model loading, and inference).
Note that each stage is implemented as an \code{ECALL} but may perform multiple \code{OCALLs}.
The measurement is repeated $30$ times and the arithmetic median is reported.

\begin{table}[t]
	\caption{The execution time (milliseconds) and overhead of the machine learning service using our system.}
	\begin{center}
		\resizebox{0.5\linewidth}{!}{%
			\begin{threeparttable}
				\begin{tabular}{|c|c|c|c|}
					\hline
					&\multicolumn{1}{|c|}{Original}&\multicolumn{1}{|c|}{\sysname}&\multicolumn{1}{|c|}{Overhead} \\
					\hline
					Enclave creation & $283.80$  & $283.91$ & $0.04\%$ \\
					\hline
					Model loading & $43.59$ & $54.00$ &  $23.90\%$ \\
					\hline
					Inference & $1,962.10$ & $1,970.85$ & $0.45\%$  \\
					\hline
					Total & $2,289.49$ & $2,308.76$ & $0.84\%$ \\
					\hline
				\end{tabular}

		\end{threeparttable}}
		\label{eval:ml}
	\end{center}
\end{table}

As shown in Table \ref{eval:ml}, our prototype introduces negligible overhead ($0.84\%$ on average).
In our prototype, extra data copies are required to transfer the \code{ECALL} parameters to the
\textit{param\_buffer}.
As mentioned above, the size of the model weights is about 34MB.
Therefore, the overhead of model loading stage is relatively high, which achieves $23.9\%$.
For a real-world machine learning inference service, one operation of the model loading
usually serves thousands of inputs or even more.
Thus the overall overhead is acceptable,
just as shown in our experiment ($0.84\%$ for $392$ inputs).

\smallskip
\noindent
\textbf{Database}\tab
We use a ported SGX version of the SQLite~\cite{sgxsqlite} database in our evaluation.
As a lightweight database engine, SQLite is widely deployed in real-world applications.
Besides, the execution of SQLite involves many file operations on the database file, thus
causing many \code{OCALLs}.
In the evaluation, we record the time cost for four common database operations, i.e.,
\textit{INSERT, SELECT, UPDATE} and \textit{DELETE}, respectively.
Each operation is performed for $10,000$ times and the SQL statements are generated
with SQLite's performance testing script (\code{tools/speedtest.tcl}).
We repeat our measurements $15$ times and report the arithmetic mean.

\begin{table}[t]
	\caption{The execution time (milliseconds) and overhead of database operations.}
	
	\begin{center}
		\resizebox{0.45\linewidth}{!}{%
			\begin{threeparttable}
				\begin{tabular}{|c|c|c|c|}
					\hline
					&\multicolumn{1}{|c|}{Original}&\multicolumn{1}{|c|}{\sysname}&\multicolumn{1}{|c|}{Overhead} \\
					\hline
					INSERT & $1,856,643$  & $1,867,392$ & $0.58\%$ \\
					\hline
					SELECT & $8,428$ & $8,629$ &  $2.38\%$ \\
					\hline
					UPDATE & $183,271$ & $186,740$ & $1.89\%$  \\
					\hline
					DELETE & $182,417$ & $182,763$  & $0.19\%$ \\
					\hline
				\end{tabular}

		\end{threeparttable}}
		\label{eval:db}
	\end{center}
\end{table}

As shown in Table \ref{eval:db}, our prototype introduces overhead for the four operations
from $0.19\%$ to $2.38\%$ with an average overhead of $1.26\%$.
The highest overhead comes from the INSERT operation, which is $2.38\%$.
We find that there are $11,0338$ \code{OCALLs} and $10,000$ \code{ECALLs} involved for the INSERT operations.
In other words, the frequency for the INSERT operation under the original SDK achieves
$13,092$ \code{OCALLs} and $1,186$ \code{ECALLs} per second
(and $12,787$ \code{OCALLs} and $1,159$ \code{ECALLs} per second under our prototype).
Under this high frequent switching between the host application and the enclave,
the performance overhead is only $2.38\%$.

\begin{table}[]
	\caption{The evaluation result of a HTTPS web server.}
	\centering
	\resizebox{0.5\linewidth}{!}{%
		\begin{threeparttable}
			\label{eval:http-eva}
			\begin{tabular}{|r|r|r|r|r|}
				\hline
				& \multicolumn{2}{c|}{Transfer Rate (kb/s)}  & \multicolumn{1}{c|}{\multirow{2}{*}{Overhead}} & \multicolumn{1}{c|}{\multirow{2}{*}{\# of OCALLs}} \\ \cline{2-3}
				& \multicolumn{1}{c|}{Original} & \multicolumn{1}{c|}{\sysname} & \multicolumn{1}{c|}{} & \multicolumn{1}{c|}{} \\ \hline
				1  & 5.94  & 5.69  & 4.21\%   & \textasciitilde $742,000$    \\ \hline
				2  & 5.93  & 5.7   & 3.88\%   & \textasciitilde $742,270$    \\ \hline
				4  & 5.9   & 5.7   & 3.39\%   & \textasciitilde $742,810$    \\ \hline
				8  & 5.93  & 5.67  & 4.38\%   & \textasciitilde $743,890$    \\ \hline
				16 & 5.92  & 5.68  & 4.05\%   & \textasciitilde $744,960$    \\ \hline
			\end{tabular}

	\end{threeparttable}}
\end{table}

\smallskip
\noindent
\textbf{A HTTPS web server}\tab 
The mbedtls-SGX~\cite{mbedtlssgx} is a ported version of the \code{mbedtls},
which can run inside the SGX and provide an SGX-enabled TLS suite for developers.
We evaluate the performance of a simple single-thread HTTPS server, which is included in the mbedtls-SGX. 
The HTTPS server serves a static web page whose size is 108 bytes. 
We use the Apache HTTP benchmarking tool~\cite{ab} to perform the evaluation.
In each test, the server will serve $1,000$ requests. 
The concurrency level, which represents the number of multiple requests to serve at a time,
is increased from 1 to 16 with different intervals. 
The transfer rate, number of \code{OCALLs}, and the overhead are collected or computed and shown in
Table~\ref{eval:http-eva}. The introduced overhead is ranging from $3.39\%$ to $4.38\%$.

\section{Discussion}

\textbf{High-level attack}\tab
\sysname focuses on eliminating the enclave-host asymmetries so that the host-enclave interaction
can be restricted to the specified interfaces.
High-level attacks~\cite{iago, sgxwolf} based on the specified interaction
interfaces (e.g., \code{ECALL/OCALL} in Intel SGX SDK~\cite{intelsdk}) are not considered.
Nevertheless, \sysname is the foundation for the defense mechanism that targets for such high-level attacks.
We regard exploring the defense space for the high-level attacks based on \sysname as our future work.

\smallskip
\noindent
\textbf{SGX2 support}\tab
Intel released the second generation SGX in its IceLake platform, called SGX2.
Compared with SGX1, SGX2 supports dynamic enclave memory management.
It provides three new run-time primitives for an enclave, including
enclave page allocation/deallocation, enclave page permission (EPCM) restriction/extension,
and dynamic thread (TCS) creation and destruction.
Currently, the implementation of our prototype is based on SGX1.
However, the design of \sysname itself is compatible with SGX2.
First, the page allocation and the dynamic threading managements in
SGX2 require the involvement of OS. So the host application can be aware of
the corresponding operations.
Second, SGX2 enables an enclave to modify the access permission of an enclave page in
the page's corresponding EPCM entry.
It is orthogonal to \sysname's access control mechanism,
which relies on page table and PKRU register to perform access permission control.

\smallskip
\noindent
\textbf{Source code change}\tab
Our prototype keeps backward-compatibility at the source code level with minor modifications.
The required modification is to remove the use of \code{user\_check} attribute.
Our experience shows that it only requires a minor engineering effort.
Besides, we argue that the use of \code{user\_check} is not recommended.
Because it is prone to the TOCTTOU attack at the design level.
Similar discussions or opinions are also expressed in other work~\cite{sgxjail}.

\section{Related Work}

\noindent
\textbf{Untrusted enclave defenses}\tab
Several mechanisms have been proposed to confine the \textit{untrusted} enclave.
One proposal is to detect the malicious actions of an enclave by monitoring its I/O behaviors~\cite{sgxexplained},
which is not practical yet. Costan et al. also proposed a solution to perform the static analysis
on the enclave code~\cite{sgxexplained}. To deal with runtime generated code, it
requires all enclaves to include a standardized static analysis framework into themselves.
This brings two security concerns, i.e., the source credibility of the static analysis framework and  the increased
TCB introduced by the framework.
SGXJail~\cite{sgxjail} confines the enclave's behaviors by residing each enclave in a separated sandbox process.
It suffers from the scalability issue, especially for multiple enclaves and multithreading scenarios.
\sysname does not have such issue because of its in-process confinement design.
Some systems~\cite{royan,occlum} apply the software fault isolation (SFI) technique
into the enclave to confine user code inside the enclave. 
However, they do not support dynamically generated code inside the enclave.
Our system does not have this limitation.

\smallskip
\noindent
\textbf{Usage of Intel MPK and SGX}\tab
Intel MPK provides a
hardware primitive to implement efficient intra-process isolation~\cite{erim, hodor, unikernel}.
ERIM~\cite{erim} leverages MPK
to provide intra-process isolation for normal applications with low performance overhead.
However, ERIM is not applicable to SGX scenarios due to the hardware isolation of SGX.
The libmpk~\cite{libmpk} system virtualizes MPK protection keys in
software so that an unlimited number of protection keys are available.
The libmpk system is orthogonal to our work and its design about protection key virtualization
can be borrowed by our system to implement the support of unlimited number of enclaves in the future.

SGX has been leveraged to protect user code and data in different scenarios. 
For instance, VC3~\cite{vc3} uses SGX to implement trustworthy data analytics in the cloud.
LightBox~\cite{LightBox} leverages SGX to provide a highly efficient solution for middleboxes
in the cloud. To facilitate the usage of SGX, different software
development kits (SDK)~\cite{intelsdk,openenclave,asylo,rustsgxsdk} have been
provided to the developers.
Moreover, several library operating systems (libOS)~\cite{haven, graphene, occlum}
have been developed to minimize the effort of legacy code refactoring for SGX.
However, all of these SDKs and libOSes have not considered the issue of untrusted enclaves.
\sysname can be integrated into them to address this issue.

\smallskip
\noindent
\textbf{Attacks towards SGX and security enhancements for SGX}\tab
Researchers have proposed different classes of attacks towards SGX.
Most of them are side-channel based attacks.
A representative type of attack is controlled-channel
attacks~\cite{controlledchannel, pagefault, cauldron}.
Another class of attacks is based on the software vulnerabilities inside the enclave~\cite{ropsgx,coinattack}.
For instance, Khandaker et al.~\cite{coinattack} systematically summarized the attack
surface of the host-enclave interaction interfaces,
from the host application's view.

Accordingly, security enhancement mechanisms~\cite{tsgx,autarky} have been introduced to
enhance the security guarantees.
To defend against controlled-channel attacks,
T-SGX~\cite{tsgx} relies on the TSX
hardware feature to suppress page faults occurred inside the enclave. 
Besides, some defense mechanisms~\cite{sgxshield,sgxbounds,mptee}
are proposed to address the vulnerabilities inside the enclave code.
These security enhancement mechanisms can collaborate with \sysname to strengthen
the mutual distrust relationship between an enclave and its host application.

\section{Conclusion}
In this paper, we propose an efficient and scalable defense mechanism to
confine an untrusted enclave's behaviors. The threats of an untrusted enclave
come from the enclave-host asymmetries, i.e., the data access asymmetry and the
control flow asymmetry. Our solution leverages two x86 hardware features,
i.e., \textit{Intel MPK and x86 single-step mode}, to break the data access and control
flow asymmetries, respectively. We have solved two practical challenges and implemented a prototype system.
The evaluation shows the efficiency of our system, with
less than $4\%$ performance overhead.

\bibliographystyle{plain}
\bibliography{./reference}

\begin{thebibliography}{10}

\bibitem{ab}
Apache http server benchmarking tool.
\newblock \url{https://httpd.apache.org/docs/2.4/programs/ab.html}, Referenced
  September, 2020.

\bibitem{asylo}
Google asylo.
\newblock \url{https://github.com/google/asylo}, Referenced August, 2020.

\bibitem{intelsdk}
Intel sgx sdk for linux.
\newblock \url{https://github.com/intel/linux-sgx}, Referenced August, 2020.

\bibitem{mbedtlssgx}
mbedtls-sgx.
\newblock \url{https://github.com/bl4ck5un/mbedtls-SGX}, Referenced July, 2020.

\bibitem{openenclave}
Microsoft open enclave sdk.
\newblock \url{https://github.com/openenclave/openenclave}, Referenced August,
  2020.

\bibitem{sgxsqlite}
Sgx-sqlite.
\newblock \url{https://github.com/yerzhan7/SGX\_SQLite}, Referenced July, 2020.

\bibitem{idash}
Anonymous.
\newblock Privacy-preserving machine learning as a service on sgx, 2019.
\newblock \url{http://www.humangenomeprivacy.org/2019/competition-tasks.html}.

\bibitem{sgxelide}
Erick Bauman, Huibo Wang, Mingwei Zhang, and Zhiqiang Lin.
\newblock Sgxelide: Enabling enclave code secrecy via self-modification.
\newblock In {\em Proceedings of the 2018 International Symposium on Code
  Generation and Optimization}, 2018.

\bibitem{haven}
Andrew Baumann, Marcus Peinado, and Galen Hunt.
\newblock Shielding applications from an untrusted cloud with haven.
\newblock In {\em Proceedings of the 11th {USENIX} Symposium on Operating
  Systems Design and Implementation}, 2014.

\bibitem{pagefault}
Jo~Van Bulck, Nico Weichbrodt, R{\"u}diger Kapitza, Frank Piessens, and Raoul
  Strackx.
\newblock Telling your secrets without page faults: Stealthy page table-based
  attacks on enclaved execution.
\newblock In {\em Proceedings of the 26th {USENIX} Security Symposium}, 2017.

\bibitem{graphene}
Chia che Tsai, Donald~E. Porter, and Mona Vij.
\newblock Graphene-sgx: A practical library {OS} for unmodified applications on
  {SGX}.
\newblock In {\em Proceedings of the 2017 {USENIX} Annual Technical
  Conference}, 2017.

\bibitem{iago}
Stephen Checkoway and Hovav Shacham.
\newblock Iago attacks: Why the system call api is a bad untrusted rpc
  interface.
\newblock In {\em Proceedings of the 18th International Conference on
  Architectural Support for Programming Languages and Operating Systems}, 2013.

\bibitem{sgxexplained}
Victor Costan and Srinivas Devadas.
\newblock Intel sgx explained.
\newblock Cryptology ePrint Archive, Report 2016/086, 2016.
\newblock \url{https://eprint.iacr.org/2016/086}.

\bibitem{LightBox}
Huayi Duan, Cong Wang, Xingliang Yuan, Yajin Zhou, Qian Wang, and Kui Ren.
\newblock Lightbox: Full-stack protected stateful middlebox at lightning speed.
\newblock In {\em Proceedings of the 26th ACM Conference on Computer and
  Communications}, 2019.

\bibitem{kernelisolation}
Daniel Gruss, Dave Hansen, and Brendan Gregg.
\newblock Kernel isolation: From an academic idea to an efficient patch for
  every computer.
\newblock 2018.

\bibitem{hodor}
Mohammad Hedayati, Spyridoula Gravani, Ethan Johnson, John Criswell, Michael~L.
  Scott, Kai Shen, and Mike Marty.
\newblock Hodor: Intra-process isolation for high-throughput data plane
  libraries.
\newblock In {\em Proceedings of the 2019 USENIX Annual Technical Conference},
  2019.

\bibitem{royan}
Tyler Hunt, Zhiting Zhu, Yuanzhong Xu, Simon Peter, and Emmett Witchel.
\newblock Ryoan: A distributed sandbox for untrusted computation on secret
  data.
\newblock In {\em Proceedings of the 12th USENIX Symposium on Operating Systems
  Design and Implementation}, 2016.

\bibitem{coinattack}
Mustakimur~Rahman Khandaker, Yueqiang Cheng, Zhi Wang, and Tao Wei.
\newblock Coin attacks: On insecurity of enclave untrusted interfaces in sgx.
\newblock In {\em Proceedings of the 25th International Conference on
  Architectural Support for Programming Languages and Operating Systems}, 2020.

\bibitem{sgxbounds}
Dmitrii Kuvaiskii, Oleksii Oleksenko, Sergei Arnautov, Bohdan Trach, Pramod
  Bhatotia, Pascal Felber, and Christof Fetzer.
\newblock Sgxbounds: Memory safety for shielded execution.
\newblock In {\em Proceedings of the 12th European Conference on Computer
  Systems}, 2017.

\bibitem{ropsgx}
Jaehyuk Lee, Jinsoo Jang, Yeongjin Jang, Nohyun Kwak, Yeseul Choi, Changho
  Choi, Taesoo Kim, Marcus Peinado, and Brent~ByungHoon Kang.
\newblock Hacking in darkness: Return-oriented programming against secure
  enclaves.
\newblock In {\em Proceedings of the 26th {USENIX} Security Symposium}, 2017.

\bibitem{sgxwolf}
Marion Marschalek.
\newblock The wolf in sgx clothing.
\newblock \url{https://www.troopers.de/troopers18/agenda/dldub8/}, Referenced
  July, 2020.

\bibitem{autarky}
Meni Orenbach, Andrew Baumann, and Mark Silberstein.
\newblock Autarky: Closing controlled channels with self-paging enclaves.
\newblock In {\em Proceedings of the 15th European Conference on Computer
  Systems}, 2020.

\bibitem{libmpk}
Soyeon Park, Sangho Lee, Wen Xu, HyunGon Moon, and Taesoo Kim.
\newblock libmpk: Software abstraction for intel memory protection keys (intel
  {MPK}).
\newblock In {\em Proceedings of the 2019 USENIX Annual Technical Conference},
  2019.

\bibitem{enclavedb}
Christian Priebe, Kapil Vaswani, and Manuel Costa.
\newblock Enclavedb: A secure database using sgx.
\newblock In {\em Proceedings of the 2018 IEEE Symposium on Security and
  Privacy}, 2018.

\bibitem{vc3}
F.~{Schuster}, M.~{Costa}, C.~{Fournet}, C.~{Gkantsidis}, M.~{Peinado},
  G.~{Mainar-Ruiz}, and M.~{Russinovich}.
\newblock Vc3: Trustworthy data analytics in the cloud using sgx.
\newblock In {\em Proceedings of the 2015 IEEE Symposium on Security and
  Privacy}, 2015.

\bibitem{sgxrop}
Michael Schwarz, Samuel Weiser, and Daniel Gruss.
\newblock Practical enclave malware with intel {SGX}.
\newblock 2019.

\bibitem{sgxshield}
Jaebaek Seo, Byoungyoung Lee, Seongmin Kim, Ming-Wei Shih, Insik Shin, Dongsu
  Han, and Taesoo Kim.
\newblock Sgx-shield: Enabling address space layout randomization for sgx
  programs.
\newblock 01 2017.

\bibitem{occlum}
Youren Shen, Hongliang Tian, Yu~Chen, Kang Chen, Runji Wang, Yi~Xu, Yubin Xia,
  and Shoumeng Yan.
\newblock Occlum: Secure and efficient multitasking inside a single enclave of
  intel sgx.
\newblock In {\em Proceedings of the 25th International Conference on
  Architectural Support for Programming Languages and Operating Systems}, 2020.

\bibitem{tsgx}
Ming-Wei Shih, Sangho Lee, Taesoo Kim, and Marcus Peinado.
\newblock T-sgx: Eradicating controlled-channel attacks against enclave
  programs.
\newblock In {\em Proceedings of the 2017 Network and Distributed System
  Security Symposium 2017}, 2017.

\bibitem{unikernel}
Mincheol Sung, Pierre Olivier, Stefan Lankes, and Binoy Ravindran.
\newblock Intra-unikernel isolation with intel memory protection keys.
\newblock In {\em Proceedings of the 16th ACM SIGPLAN/SIGOPS International
  Conference on Virtual Execution Environments}, 2020.

\bibitem{erim}
Anjo Vahldiek-Oberwagner, Eslam Elnikety, Nuno~O. Duarte, Michael Sammler,
  Peter Druschel, and Deepak Garg.
\newblock {ERIM}: Secure, efficient in-process isolation with protection keys
  ({MPK}).
\newblock In {\em Proceedings of the 28th USENIX Security Symposium}, 2019.

\bibitem{teeassess}
Jo~Van~Bulck, David Oswald, Eduard Marin, Abdulla Aldoseri, Flavio~D. Garcia,
  and Frank Piessens.
\newblock A tale of two worlds: Assessing the vulnerability of enclave
  shielding runtimes.
\newblock In {\em Proceedings of the 2019 ACM SIGSAC Conference on Computer and
  Communications Security}, 2019.

\bibitem{rustsgxsdk}
Huibo Wang, Pei Wang, Yu~Ding, Mingshen Sun, Yiming Jing, Ran Duan, Long Li,
  Yulong Zhang, Tao Wei, and Zhiqiang Lin.
\newblock Towards memory safe enclave programming with rust-sgx.
\newblock In {\em Proceedings of the 2019 ACM SIGSAC Conference on Computer and
  Communications Security}, 2019.

\bibitem{cauldron}
Wenhao Wang, Guoxing Chen, Xiaorui Pan, Yinqian Zhang, XiaoFeng Wang, Vincent
  Bindschaedler, Haixu Tang, and Carl~A. Gunter.
\newblock Leaky cauldron on the dark land: Understanding memory side-channel
  hazards in sgx.
\newblock In {\em Proceedings of the 2017 ACM SIGSAC Conference on Computer and
  Communications Security}, 2017.

\bibitem{sgxjail}
Samuel Weiser, Luca Mayr, Michael Schwarz, and Daniel Gruss.
\newblock Sgxjail: Defeating enclave malware via confinement.
\newblock In {\em Proceedings of the 22nd International Symposium on Research
  in Attacks, Intrusions and Defenses}, 2019.

\bibitem{controlledchannel}
Yuanzhong Xu, Weidong Cui, and Marcus Peinado.
\newblock Controlled-channel attacks: Deterministic side channels for untrusted
  operating systems.
\newblock In {\em Proceedings of the 2015 IEEE Symposium on Security and
  Privacy}, 2015.

\bibitem{mptee}
Wenjia Zhao, Kangjie Lu, Yong Qi, and Saiyu Qi.
\newblock Mptee: Bringing flexible and efficient memory protection to intel
  sgx.
\newblock In {\em Proceedings of the 15th European Conference on Computer
  Systems}, 2020.

\end{thebibliography}

\end{document}